\documentclass[pdflatex,sn-mathphys-num]{sn-jnl}

\usepackage[none]{hyphenat}
\usepackage{geometry}
\usepackage{graphicx}%
\usepackage{multirow}%
\usepackage{amsmath,amssymb,amsfonts}%
\usepackage{amsthm}%
\usepackage[title]{appendix}%
\usepackage{xcolor}%
\usepackage{textcomp}%
\usepackage{manyfoot}%
\usepackage{booktabs}%
\usepackage{algorithm}%
\usepackage{algorithmicx}%
\usepackage{algpseudocode}%
\usepackage{listings}%
\usepackage{fancyvrb}
\usepackage{changepage} 
\usepackage[most]{tcolorbox}
\tcbuselibrary{listingsutf8} 
\usepackage{adjustbox}
\usepackage[most]{tcolorbox}
\usepackage{enumitem}
\usepackage{multirow}
\usepackage{float}
\usepackage[most]{tcolorbox}
\usepackage{enumitem}
\usepackage{tabularx} 
\usepackage{lmodern}
\usepackage[T1]{fontenc}

\theoremstyle{thmstyleone}%
\theoremstyle{thmstyletwo}%

\theoremstyle{thmstylethree}%

\begin{document}

\title[Article Title]{Interpretability Framework for LLMs in Undergraduate Calculus}


\author[1]{\fnm{Sagnik} \sur{Dakshit}}\email{sdakshit@uttyler.edu}

\author[2]{\fnm{Sushmita} \sur{Sinha Roy}}\email{ssinharoy@fgcu.edu}

\affil[1]{\orgdiv{Computer Science}, \orgname{University of Texas at Tyler}, \orgaddress{\street{3900 University Blvd}, \city{Tyler}, \postcode{75799}, \state{TX}, \country{USA}}}

\affil[2]{\orgdiv{Department of Mathematics}, \orgname{Florida Gulf Coast University}, \orgaddress{\street{10501 FGCU Blvd}, \city{Fort Myers}, \postcode{33965}, \state{FL}, \country{USA}}}


\abstract{
Large Language Models (LLMs) are increasingly being used in education, yet their correctness alone does not capture the quality, reliability, or pedagogical validity of their problem-solving behavior, especially in mathematics, where multistep logic, symbolic reasoning, and conceptual clarity are critical. Conventional evaluation methods largely focus on final answer accuracy and overlook the reasoning process. To address this gap, we introduce a novel interpretability framework for analyzing LLM-generated solutions using undergraduate calculus problems as a representative domain. Our approach combines reasoning flow extraction and decomposing solutions into semantically labeled operations and concepts with prompt ablation analysis to assess input salience and output stability. Using structured metrics such as reasoning complexity, phrase sensitivity, and robustness, we evaluated the model behavior on real Calculus I–III university exams. Our findings revealed that LLMs often produce syntactically fluent yet conceptually flawed solutions, with reasoning patterns sensitive to prompt phrasing and input variation. This framework enables fine-grained diagnosis of reasoning failures, supports curriculum alignment, and informs the design of interpretable AI-assisted feedback tools. This is the first study to offer a structured, quantitative, and pedagogically grounded framework for interpreting LLM reasoning in mathematics education, laying the foundation for the transparent and responsible deployment of AI in STEM learning environments.
}




\maketitle

\section{Introduction}

Large Language Models (LLMs) have demonstrated remarkable capabilities in natural language understanding, question answering, and even basic symbolic reasoning. As these systems become increasingly integrated into educational environments ranging from intelligent tutoring systems to AI-assisted assessment, the need to evaluate and interpret problem-solving behavior has grown urgent, especially in structured and logic-intensive domains like mathematics. Evaluating whether these systems truly “understand” mathematical reasoning remains an open question, especially when it comes to multi-step logic, symbolic manipulation, and conceptual abstraction. Traditional evaluation methods that focus solely on answer correctness often obscure how an LLM arrives at its solution or whether the reasoning process is pedagogically sound.

While prior studies have explored LLMs' performance on factual or conceptual language tasks, significantly less attention has been paid to how these models approach multistep mathematical reasoning. Mathematics, particularly calculus, demands precision, symbolic manipulation, and logical structural characteristics that go beyond surface-level fluency. Merely assessing whether a model generates the correct answer is insufficient to understand the mechanisms or rationale behind its conclusions, nor does it evaluate the internal consistency of its problem-solving process. For large language models (LLMs) to be responsibly integrated into educational environments, particularly within STEM fields, it is imperative to extend beyond mere accuracy and develop methodologies to systematically interpret reasoning processes. To address this gap, we introduce a structured interpretability framework designed to analyze and explain how LLMs reason through mathematical problems. Our framework consists of two primary components: (1) a reasoning flow analysis that breaks down model outputs into labeled operations, concepts, and complexity scores, and (2) a prompt sensitivity ablation method that quantifies the influence of specific input elements on output behavior. Our interpretability framework contributes to the first post-hoc decomposition pipeline for LLM-generated mathematical solutions that combines structural reasoning flow and ablation analysis, as illustrated in Section \ref{sec:Framework}. By reconstructing how models reason or misreason and identifying which input elements most affect that reasoning, we provide tools for:
\begin{itemize}
    \item \textbf{Curriculum Alignment:} Evaluate whether model reasoning follows pedagogically valid patterns.
    \item \textbf{Failure Prediction:} Flag brittle responses to complex or under specified prompts.
    \item \textbf{Retrieval Tuning:} Quantify how retrieval strategies affect robustness of generated solutions.
    \item \textbf{Training Prioritization:} Identify operations and concepts where models consistently struggle.
\end{itemize}

\noindent This enable deeper insight into model reasoning, beyond accuracy metrics. We apply this framework in a real-world case study evaluating the behavior of LLMs on Calculus I–III university exams and becnhmarking against real students’ scores. These exams serve as rigorous, structured testbeds, allowing us to examine how well the model handles varying levels of mathematical complexity and abstraction. While we report standard performance scores against student average scores graded by mathematical faculty experts, the primary goal of this study is to demonstrate the usefulness of our framework in identifying reasoning patterns, inconsistencies, and prompt sensitivities that would otherwise remain hidden while benchmarking the performance and not comprehensively evaluating all LLM model performance on calculus. We further tested the performance of LLMs by incorporating external knowledge using retrieval augmented generation (RAG) and contextual retrieval methods, evaluating how these affect not only accuracy but also reasoning fidelity. Contrary to expectations, naive retrieval strategies often degrade performance, underscoring the importance of context alignment and interpretability when augmenting LLMs in high-stakes domains. Additionally, we discuss expert feedback on the limitations of LLM performance in calculus problem-solving. 

\subsection{Motivation: Need for Reasoning Centered Interpretability}
\label{sec:motivation}

While large language models (LLMs) are increasingly capable of solving mathematical problems, their deployment in real-world education remains constrained by a lack of transparent reasoning. In high-stakes STEM domains, correctness alone is insufficient; both educators and learners must understand the \textit{how} and \textit{why} behind a model’s solution. LLM-generated responses can be superficially fluent yet pedagogically misleading, making it difficult to assess whether reasoning is valid, shallow, or inconsistent.

This gap motivates the need for interpretability frameworks tailored to education. Standard metrics such as rubric scores or end-to-end accuracy fail to capture multistep logic, symbolic transformations, or sensitivity to prompt phrasing. To support instructional alignment and responsible model use, educators require tools to trace reasoning failures, uncover conceptual misunderstandings, and identify brittle dependencies in model behavior. Our work addresses this by proposing a post-hoc interpretability framework that decomposes LLM solutions into reasoning chains, evaluates robustness through controlled perturbations, and quantifies step complexity, sensitivity, and stability. These methods offer actionable insights for debugging models, aligning AI outputs with curricula, and enhancing learner trust.

\subsection{Calculus as a Case Study}
\label{sec:calculus}

Calculus provides an ideal domain for evaluating reasoning in LLMs due to its structured progression of complexity and central role in undergraduate STEM education. Most university curricula divide calculus into three sequential courses Calculus I, II, and III—typically completed during the first two years of study. These courses not only build mathematical fluency but also serve as a foundation for advanced topics in engineering, physics, computer science, and economics. Calculus I introduces core concepts such as limits, continuity, derivatives, and introductory integration, emphasizing procedural fluency and practical problem solving. Calculus II deepens abstraction through integration techniques, series, and parametric representations, requiring logical reasoning and conceptual synthesis. Calculus III advances to multivariable functions, vector calculus, and 3D geometric modeling, demanding spatial reasoning and symbolic precision.
This progression from concrete computation to abstract generalization—offers a natural framework for probing LLM behavior across increasing cognitive demands. By applying our interpretability framework to real Calculus I–III exams, we can assess how LLMs handle procedural steps, symbolic manipulations, and conceptual understanding across a range of mathematical tasks.

\subsection{Contributions}

This paper makes the following contributions to the fields of educational AI and explainable machine learning:

\begin{enumerate}
    \item \textbf{Interpretability framework for LLMs in mathematics:} We introduce a domain-grounded interpretability framework that decomposes LLM-generated solutions into semantically annotated reasoning steps and evaluates sensitivity to prompt phrasing through controlled ablations.

    \item \textbf{Structured metrics for reasoning analysis:} We define and operationalize quantitative metrics such as reasoning complexity, robustness, phrase sensitivity, and step count that enable fine-grained evaluation of model behavior beyond correctness.

    \item \textbf{Empirical case study using real calculus exams:} We apply the framework to a series of Calculus I–III university exams, benchmarking LLM performance against human student scores and examining the effects of knowledge augmented configurations.

    \item \textbf{Insights into pedagogical alignment and model limitations:} Our analysis reveals common reasoning failures, sensitivity to linguistic variation, and gaps in conceptual understanding, offering guidance for responsible classroom use and instructional alignment.
\end{enumerate}

\noindent This is the first study to focus on interpreting and analyzing LLM reasoning in calculus problem solving through reasoning flow and ablation studies, along with benchmarking against real-world student performance.  By making reasoning processes transparent, our framework supports the design of interpretable and accountable AI tools for use in formative feedback, grading, and tutoring in STEM education.

The rest of the paper is organized as follows: Section \ref{sec:litsurvey}  discusses related research. Section \ref{sec:motivation} provides detailed motivation for the need for an interpretable framework in LLM reasoning and our choice of calculus as a use case. Section \ref{sec:Framework} explains the proposed framework and its evaluation. Section \ref{sec:setup} elaborates our experimental methodology, and Section \ref{sec:results} presents the experimental results.
Finally, Section \ref{sec:expertfeedback} presents mathematical expert observations in LLM prowess in calculus, pedagogical implications, and future scope to address limitations and lastly in Section \ref{sec:conclusion} we present concluding remarks of this
paper.

\section{Related Works}
\label{sec:litsurvey}

\subsection{Explainable AI in Education}

Recent advancements in educational AI emphasize the growing importance of interpretability, fairness, and human-centered design, particularly in the deployment of large language models (LLMs) in classroom settings. Prior work by Maity et al. \cite{maity2024humancentricexplainableaieducation} underscores the value of explanations that are not only technically accurate but also pedagogically meaningful, advocating for transparency and cultural sensitivity in AI-generated feedback. In alignment with this vision, our work focuses on making LLM outputs interpretable in a manner that supports student learning and instructor insight into calculus. Chinta et al. \cite{chinta2024fairaiednavigatingfairnessbias} explored fairness-aware design in AI for education, noting the potential for algorithmic bias and the necessity of stakeholder-informed development. Their findings motivate our focus on fine-grained reasoning analysis, which reveals disparities in how LLMs handle procedural versus conceptual tasks, an important consideration for equitable learning support. Schneider et al. \cite{schneider2024explainablegenerativeaigenxai} provided a comprehensive review of explainability in generative AI, including interactivity, verifiability, and audience-specific customization. While these principles have seen limited adoption in educational contexts, our framework operationalizes them by decomposing mathematical reasoning into annotated steps, revealing both the strengths and breakdowns of model logic. Concerns about generative AI alignment in education are further highlighted by Gupta et al. \cite{gupta2024generative}, who noted risks such as hallucinations, curriculum misalignment, and lack of cultural grounding. These issues are particularly acute in mathematics, where precision and conceptual fidelity are critical. Our interpretability framework responds by exposing subtle flaws and missteps in LLM outputs, even when the final answers appear correct. 

From a policy standpoint, the U.S. Department of Education 
\footnote{\url{https://www.ed.gov/sites/ed/files/documents/ai-report/ai-report.pdf}}, outlines principles for responsible AI in education, calling for contextual adaptation, iterative evaluation, and support for learner agency. Our work echoes this by offering a systematic, data-driven method for evaluating LLM behavior in real curricular scenarios, enabling continuous refinement and oversight. Ethical considerations have also been emphasized by Memarian et al. \cite{memarian2023fairness}, who advocate participatory design and stakeholder involvement in AI systems used in higher education. Our approach supports this by providing interpretable artifacts (e.g., reasoning graphs and sensitivity scores) that can be meaningfully used by students and instructors alike. Our study addresses these concerns by presenting structured quantitative metrics tailored to mathematics tasks, grounded in domain expertise, and validated through real exam questions.

Foundational taxonomies of interpretability methods, such as those presented by \cite{8466590}, help situate our method along the spectrum of post-hoc explanation tools. By blending symbolic reasoning patterns with semantic annotations, our framework offers a hybrid approach that balances fidelity and usability for both technical and non-technical users. Wang et al. \cite{wang2024large} propose LLM based symbolic programs as a step toward more transparent AI systems, combining neural outputs with rule-based logic. While our work does not rely on symbolic synthesis, it shares a similar motivation for making the internal reasoning of LLMs observable and aligned with human expectations in educational contexts. The theoretical contributions of this study also shape our approach. Rosé et al. \cite{rose2019explanatory} stressed the need for AI explanations that support sensemaking and metacognition rather than merely improving prediction accuracy. Clancey \cite{clancey2021methods} extends this argument by framing explanation as a socially situated process that is dynamic, dialogic, and deeply dependent on context. These ideas emphasize step-level analysis and prompt sensitivity as a means of reconstructing how the model understands and solves mathematical problems. Finally, broader surveys in explainable ML \cite{escalante2018explainable, linardatos2020explainable} identify key metrics of explanatory quality, such as cognitive load, user trust, and task relevance. Our interpretability framework is informed by these principles, prioritizing explanations that are not only technically valid but also cognitively useful in high-stakes, domain-specific learning environments such as undergraduate calculus.

Collectively, these studies provide a robust foundation for evaluating and designing explainable AI systems in education. However, few studies have rigorously examined LLM reasoning on domain-specific assessments at a fine-grained level. Our study addresses this gap by offering a comprehensive interpretability framework that captures reasoning structure, input sensitivity, and step complexity, furnishing insights beyond accuracy that are essential for responsible educational deployment.

\subsection{Reasoning Chains in NLP}

Reasoning chains in natural language processing (NLP) is pivotal to enhancing machine understanding and response generation capabilities. They represent the sequences of logical steps or decisions that an AI model takes to arrive at a conclusion. Within the context of NLP, reasoning chains can significantly impact tasks such as question answering, natural language inference, and multi-hop reasoning, where models are required to connect disparate pieces of information to generate coherent and relevant outputs. The development of reasoning chains involves advanced techniques such as Chain-of-Thought (CoT) prompting, which encourages models to generate intermediate reasoning steps before producing a final answer. This method aims to enhance the comprehensiveness and accuracy of the responses generated by language models. For instance, the LLaMA-LoRA model combines neural prompt engineering with low-rank adaptation to improve logical reasoning capabilities in Chinese NLP, surpassing benchmarks set by models such as GPT-3.5 \cite{chen2024llama}. Despite advancements, challenges remain, as seen in models such as ChatGPT, which excels in certain reasoning tasks but struggles with others, such as commonsense reasoning and summarization. This inconsistency underscores the need for continued evolution in AI models to effectively handle diverse NLP tasks \cite{wu2024evaluating}. An extensive survey of natural language reasoning emphasizes the importance of both classical logical reasoning and emerging avenues, such as defeasible reasoning in NLP. The focus is on constructing precise reasoning frameworks that incorporate external knowledge, thereby advancing the state of AI-driven language models \cite{yu2024natural}. Moreover, natural language inference (NLI) tasks have benefited from utilizing external knowledge bases, highlighting the potential of structured knowledge in enhancing model performance in specific domains, such as science questions\cite{wang2019improving}. In summary, reasoning chains in NLP are essential for achieving higher-order logic and improving model comprehension. Continued research and refinement in this area promises to address current limitations, guiding the next generation of AI language models toward more robust and nuanced understanding and reasoning capabilities.

\subsection{LLMs in Mathematics}

The reasoning capabilities of Large Language Models (LLMs) in mathematics have garnered substantial interest owing to the inherent complexities involved in mathematical problem solving. LLMs, such as GPT-3 and GPT-4, demonstrate considerable potential in natural language processing tasks and have extended their capabilities to solving certain reasoning problems in mathematics. However, despite their impressive performance, several aspects underline their limitations in terms of mathematical reasoning. One primary limitation is that LLMs often display shallow reasoning capabilities, even with advanced prompting techniques. Formal logic, which is essential for complex reasoning, remains a challenge for LLMs, as they struggle to accurately convert natural language into formal logic. One study highlighted the potential of LLMs in transforming natural language descriptions of logic puzzles into answer set programs, demonstrating that with carefully designed prompts and few learning examples, LLMs could assist in creating complex logic programs. However, errors are common and simple enough for human intervention \cite{ishay2023leveraging}. LLMs have been subjected to tests such as GSM8K to evaluate their mathematical reasoning. Although their performance has improved, studies indicate that these improvements do not necessarily translate into genuine advancements in reasoning capabilities. The models often replicate reasoning steps from their training data rather than performing genuine logical reasoning \cite{mirzadeh2024gsm}. Furthermore, a dynamically adaptive framework has been proposed to classify errors in LLMs' approach to Math Word Problems (MWPs). Traditional error classification is limited by static categories, thus failing to accurately reflect the different error patterns in mathematical reasoning. By analyzing 15 LLMs across several datasets, researchers introduced MWPES-300K, a dataset designed for comprehensive error analysis. This study showed that as models become more sophisticated, their error patterns shift from basic to complex, emphasizing the need for error-aware prompting strategies to improve mathematical reasoning performance \cite{sun2025error}. Moreover, a distinct prompting strategy termed the Pedagogical Chain-of-Thought was devised to enhance LLMs' ability to detect reasoning mistakes, particularly in mathematics. By employing educational theory principles, this approach reportedly outperforms traditional prompting strategies in identifying errors, thus laying the groundwork for more reliable math-answer grading \cite{jiang2024llms}. Fine-tuning LLMs for mathematical reasoning have shown promise. MetaMath, a specialized model trained on a dataset named MetaMathQA, has demonstrated superior performance on mathematical reasoning benchmarks compared to many open-source models. It outperformed others in the GSM8K and MATH benchmarks by a significant margin, showcasing the potential of targeted fine-tuning in enhancing mathematical reasoning capabilities \cite{yu2023metamath}.

Recent literature also highlights the growing role of Large Language Models (LLMs) in supporting mathematics education, particularly in domains such as calculus. Liu and Yang \cite{liu2024application} examine how LLMs assist students in programming related tasks by providing guidance in mathematical logic and reasoning, thereby enhancing conceptual understanding in calculus. In addition to computational support, these models help students engage more deeply with problem-solving processes, especially in engineering contexts where tools such as MATLAB are commonly used. Sharma et al. \cite{sharma2025role} explore the capacity of LLMs to deliver personalized learning experiences. By adapting responses to individual queries, these systems help tailor instruction to student-specific needs, improving engagement and academic outcomes across a variety of STEM disciplines, including calculus. This personalized support positions LLMs as adaptive educational tools capable of complementing traditional instruction. Bernabei et al. \cite{bernabei2023students} document the increasing use of LLMs, such as ChatGPT, by students for content generation, academic assessment preparation, and mathematical problem solving. These tools act as virtual tutors or writing assistants, enabling students to refine their understanding and articulate solutions more effectively. Despite their promise, the integration of LLMs into educational workflows is not without challenges. Kang et al. \cite{kang2024integrating} and Wang et al. \cite{wang2024largeEd} raise concerns regarding the accuracy and ethical implications of AI generated content. They stress the importance of validating AI outputs and ensuring that LLM-assisted learning preserves academic integrity. These studies collectively call for hybrid educational approaches that combine the efficiency of AI with pedagogically sound, human-guided learning frameworks.

Although existing research has made significant strides in understanding the potential of LLMs and explainable AI in educational settings, several key limitations persist. Most prior studies have focused on high-level applications or generic instructional benefits, without delving into domain-specific evaluations of reasoning quality or interpretability at a granular level. Furthermore, few studies offer systematic frameworks for identifying latent errors, evaluating sensitivity to prompt changes, or decomposing mathematical reasoning into meaningful pedagogical components. Our study addresses these gaps by introducing a domain-grounded interpretability framework tailored to undergraduate calculus assessments. Unlike prior work in explainable AI for education, which typically focuses on language tasks or limited numerical feedback, our framework provides symbolic traceability, sensitivity quantification, and cognitive complexity estimation metrics rarely explored in the context of undergraduate mathematics education. By combining symbolic annotation, reasoning decomposition, and robustness analysis, we move beyond accuracy to illuminate how, why, and when LLMs succeed or fail in an essential step toward safe, transparent, and educationally effective AI deployment in mathematics education.

\section{Interpretability Framework and Quantitative Evaluation}
\label{sec:Framework}

\begin{figure}[htbp]
    \centering
    \includegraphics[width=0.95\textwidth]{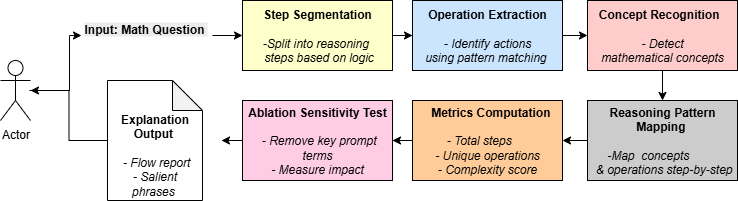}
    \caption{Flow Reasoning Framework: A visual overview of the step wise process used to analyze LLM generated responses, including operation and concept extraction, reasoning pattern detection, and ablation based prompt sensitivity analysis.}
    \label{fig:flow-reasoning-framework}
\end{figure}

To move beyond correctness and enable deeper insight into LLM behavior on mathematical tasks, in this paper, we introduce a structured interpretability framework that analyzes both the reasoning trajectories and input dependencies of LLM-generated solutions, as presented in Fig. \ref{fig:flow-reasoning-framework}. Our framework comprises two primary complementary modules: (1) reasoning flow module, which decomposes step-by-step outputs into semantically annotated chains of thought, and (2) sensitivity module, which quantifies the influence of input components through controlled perturbation. We further grounded our framework components in quantitative metrics, enabling fine-grained comparisons across questions, systems, and retrieval strategies.

\subsection{Structured Reasoning Flow Analysis}
\label{sec:flow}
Our reasoning flow module presents a systematic method for extracting and analyzing the reasoning flows embedded in LLM-generated calculus solutions. Starting from detailed step-by-step outputs produced by the model, we segment these solutions into discrete reasoning steps using rule-based parsing. Each step is then semantically annotated along the three critical dimensions as follows:

\begin{itemize}
\item \textbf{Mathematical Operation:} Fundamental actions such as differentiation, substitution, simplification, or integration.
\item \textbf{Conceptual Tag:} Domain specific concepts including the chain rule, directional derivatives, or the Fundamental Theorem of Calculus.
\item \textbf{Cognitive Complexity:} An estimate of abstraction or difficulty level (low, medium, high), derived from symbolic depth and syntactic complexity.
\end{itemize}

These annotated steps form a directed reasoning graph, where nodes represent individual reasoning steps, and edges capture logical dependencies. From this structure, we compute summary statistics, such as the total operation count, unique concepts used, average complexity per step, and a composite reasoning complexity score. This approach provides a transparent and interpretable reconstruction of the model’s latent reasoning process, enabling the detection of skipped justifications, conceptual errors, or over-generalizations that are not evident from final correctness alone.

\subsection{Sensitivity Ablation and Input Salience}
\label{sec:ablation}

Our sensitivity module systematically quantifies the sensitivity of LLM outputs to specific input question components. We implemented a structured ablation procedure, followed by divergence scoring, to quantify the impact score. Each calculus question prompt passed to the LLM is decomposed into constituent elements, including

\begin{itemize}
    \item \textbf{Mathematical Operation} (e.g., $f(x)$, $\nabla f$, coordinate vectors),
    \item \textbf{Instructional keywords} (e.g., ``find'', ``compute'', ``evaluate''),
    \item \textbf{Numerical and coordinate references} (e.g., $(x, y) = (-1, 4)$),
    \item \textbf{Linguistic features} (e.g., punctuation, casing, or word order).
\end{itemize}

For each ablation trial, one element was masked or syntactically perturbed while preserving the rest of the prompt. For example, removing the reference $(-1, 4)$ from the prompt ``Find the gradient $\nabla f$ of $f(x, y) = x^2y$ at point $(-1, 4)$'' results in the modified input: ``Find the gradient $\nabla f$ of $f(x, y) = x^2y$''. The LLM is queried with both the original and ablated prompts. To evaluate the semantic and structural divergence between the original output and each ablated output, we employ

\begin{itemize}
    \item \textbf{TF-IDF Weighted Cosine Similarity}: Outputs are vectorized using TF-IDF representations, and cosine similarity is computed to assess semantic shifts in topic and terminology.
    \item \textbf{Normalized Edit Distance (Levenshtein Distance)}: Measures token-level changes between original and ablated responses, reflecting structural or syntactic perturbations.
\end{itemize}

These metrics were aggregated across all ablations for a given question to compute two quantized interpretability scores:

\begin{itemize}
    \item \textbf{Phrase Sensitivity}: The maximum divergence observed across all ablations, reflecting the most influential input token or phrase for each question.
    \item \textbf{Robustness Score}: The average semantic similarity across ablated responses, capturing the overall output stability for each question.
\end{itemize}

This analysis reveals how sensitive the model’s reasoning is to specific elements in the prompt, identifying brittle linguistic dependencies, and conceptually salient input features. Our approach allows educators and researchers to understand which parts of a problem influence model behavior the most, and where reasoning stability may break down.

\subsection{Quantitative Metrics and Aggregation}
\label{sec:metrics}
This section elaborates our rigorous approach for evaluating the proposed interpretability framework by computing the quantitative metrics at the question level that capture key aspects of model behavior, such as semantic stability, reasoning complexity, reasoning pattern details, and sensitivity to phrasing variations. These detailed question-level metrics offer insights into how the model processes individual problems. Subsequently, these metrics are then aggregated to produce exam-level summaries and presented in \ref{sec:results} for the dual purposes of 1) characterizing overall LLM performance and reasoning patterns across the full assessment and 2) clarity of presentation of results in this paper. This section first defines and interprets each question-level metric, followed by aggregated exam-level metrics, concluding with a discussion of the importance of these metrics for interpreting large language model outputs in mathematical problem-solving tasks.

\begin{tcolorbox}[
  colback=gray!5!white,
  colframe=black!50,
  title=Question Level Metrics: Interpretation,
  fonttitle=\bfseries,
  sharp corners,
  boxrule=0.5pt,
  left=0pt,
  right=0pt,
  top=2pt,
  bottom=2pt,
  before skip=5pt,
  after skip=5pt,
  enhanced,
  attach boxed title to top center={yshift=-2mm},
  boxed title style={size=small,colback=gray!75!white}
]
\renewcommand{\arraystretch}{1.2}
\begin{tabularx}{\textwidth}{|l|X|}
\hline
\textbf{Metric} & \textbf{Interpretation} \\
\hline
\textbf{Robustness} & Measures model stability to variations in question phrasing. \\
\hline
\textbf{Impact Factor} & Identifies tokens or critical elements most influential to the model's output. \\
\hline
\textbf{Step Count} & Reflects the depth and complexity of the model's multi-step reasoning. \\
\hline
\textbf{Complexity} & Quantifies the difficulty and reasoning demands of the question. \\
\hline
\textbf{Reasoning Pattern Trace} & Captures the model's reasoning approach and patterns through the question. \\
\hline
\textbf{Phrase Sensitivity} & Measures how sensitive the model's response is to changes in phrasing. \\
\hline
\end{tabularx}
\end{tcolorbox}

\begin{tcolorbox}[
  colback=gray!5!white,
  colframe=black!50,
  title=Exam-Level Interpretation and Evaluation Metrics,
  fonttitle=\bfseries,
  sharp corners,
  boxrule=0.5pt,
  left=0pt,
  right=0pt,
  top=2pt,
  bottom=2pt,
  before skip=5pt,
  after skip=5pt,
  enhanced,
  attach boxed title to top center={yshift=-2mm},
  boxed title style={size=small,colback=gray!75!white}
]
\renewcommand{\arraystretch}{1.2}
\begin{tabularx}{\textwidth}{|l|X|}
\hline
\textbf{Metric} & \textbf{Notes} \\
\hline
\textbf{Robustness} & Measures model stability to phrasing variations across all questions in an exam. \\
\hline
\textbf{Complexity} & Based on individual question features (e.g., steps, nested logic). \\
\hline
\textbf{Step Count} & Inferred from reasoning path per question, then averaged. \\
\hline
\textbf{Phrase Sensitivity} & Measures change in output with paraphrasing, per question. \\
\hline
\end{tabularx}
\end{tcolorbox}

\begin{itemize}
    \item \textbf{Robustness} measures the model’s consistency in producing correct answers across variations in input phrasing or question structure. A robust LLM maintains its performance even when questions are paraphrased or perturbed. High robustness indicates stability and reliability in real-world or noisy user settings. The standard deviation helps assess performance volatility across questions, which are then aggregated at exam-levels.

    \item \textbf{Impact Factor} identify which parts of the input question most strongly influence the LLM’s response, derived from ablation or attention analysis. By pinpointing key tokens or phrases that the model relies upon, researchers can understand the model’s focus areas and whether it attends to relevant information pertaining to mathematical reasoning. This insight helps interpret how the model reasons and can highlight potential biases or errors in understanding, improving transparency at the question level.

    \item \textbf{Complexity} evaluates the difficulty of the reasoning involved, considering factors such as the number of mathematical operations, nested logic, and conceptual constructs required. This metric contextualizes the model’s performance by linking reasoning demands to output quality at the question level, which we also aggregate at the exam level to understand the complexity of the overall exam. Understanding complexity allows researchers to identify which question types challenge the model and where interpretability methods should focus to better explain intricate reasoning paths.

    \item \textbf{Reasoning Pattern Trace} captures the sequence and types of logical or mathematical operations the LLM applies during problem solving. It provides a structured map of the model’s thought process, revealing whether it uses appropriate methods such as substitution, differentiation, or evaluation at each step for each question. This trace is crucial for interpretability because it allows researchers to verify the correctness and coherence of the model’s reasoning flow, allowing debugging and refinement.

    \item \textbf{Complexity} captures the semantic and syntactic difficulty of the questions, based on token length, number of operations, or concept depth. This metric contextualizes performance, and higher complexity correlates with increased cognitive load. Comparing performance across complexity levels helps diagnose whether models struggle disproportionately with harder questions. We aggregated the complexity results per question and assigned the overall exam a complexity score.

    \item \textbf{Step Count} represents the average number of reasoning or computational steps taken (or inferred) by the model in its answer for each question and aggregated at exam-level. Higher step counts suggest multi-hop reasoning or deeper logical chaining. This metric evaluates not only correctness, but also how the model arrives at an answer that is important for interpretability and alignment with human-like problem solving.

    \item \textbf{Phrase Sensitivity} measures the extent to which minor changes in wording or phrasing impact the model’s output. Low phrase sensitivity implies semantic understanding beyond surface patterns. High phrase sensitivity indicates an overreliance on specific phrasing or keywords, which undermines generalization. This metric is essential for gauging robustness to natural language variability for each question and then aggregated at the exam level.
    
\end{itemize}

\noindent We would like to acknowledge that no statistical testing between the student and LLM performances were conducted as we did not access individual student data but only the class averages which restricts a statistical analysis. We believe while statistical analysis are important for understand the relation between student performance and LLM performance, this is beyond the scope of this paper with the main contribution of interpretability framework and its evaluation. Additionally, under acceptance of this paper, link to our framework implementation in python will be provided for reproducible use.

\section{Experimental Setup}
\label{sec:setup}
 In this section, we discuss our experimental setup to evaluate our interpretable framework by selecting a high-performing LLM model and establishing the baseline by comparing the model performance to real student scores across multiple levels of calculus for a semester worth of exams. We also illustrate our knowledge augmentation experiments to evaluate the effect on LLM performance, namely Retrieval Augmented Generation (RAG) and Contextual Retrieval methods. Furthermore, all three experimental configurations are compared in terms of the aggregated metrics introduced in Section \ref{sec:metrics}.

 \subsection{Real World Data}
 \label{sec:classdetails}
 For benchmarking of LLM scores against real world student performance, aggregated class average for three courses namely Calculus I, II, and III were used from a US institution. Each course has three exams termed here as Exam I, II, and III following a structured curriculum. Each course has approximately 50 students and the scores presented here are average of the student population. Further details of the courses (such as course numbers, university names) are not disclosed to ensure privacy and protection of undergraduate students and faculty members. Additionally, further details of the course are not relevant to our study in evaluating our proposed framework for interpretation mathematical reasoning of LLMs in Calculus.

\subsection{Grading Rubric}
\label{sec:grading_rubric}
\noindent The following standardized rubric is used to evaluate student and LLM responses to individual exam questions in Calculus I, II, and III. Each question is graded not only on the correctness of the final answer but also on the process, clarity, and mathematical communication.

Partial credit is awarded based on the level of understanding demonstrated in the student's or LLM's work. If student or LLM applies an incorrect method but shows some relevant and conceptually related steps, it may receive up to 25--30\% of the total points for that question. If the correct method is chosen but significant errors occur in computation, algebra, calculus, or writing proper mathematical notation, typically 40--60\% of the points may be awarded, depending on the severity of the mistakes. For minor arithmetic, notation, or sign errors with an otherwise correct method and steps, partial credit of 80--90\% is usually given. Full credit is reserved for responses that are mathematically correct, logically organized, and clearly presented. Responses that are completely incorrect, incoherent, or left blank typically receive no credit.
\vspace{1em}
\begin{tcolorbox}[
  colback=gray!5!white,
  colframe=black!50,
  title=Calculus Exam Grading Rubric (Per Question),
  fonttitle=\bfseries,
  sharp corners,
  boxrule=0.5pt,
  left=0pt,
  right=0pt,
  top=2pt,
  bottom=2pt,
  before skip=5pt,
  after skip=5pt,
  enhanced,
  attach boxed title to top center={yshift=-2mm},
  boxed title style={size=small,colback=gray!75!white}
]
\begin{tabularx}{\textwidth}{|>{\raggedright\arraybackslash}X|c|>{\raggedright\arraybackslash}X|}
\hline
\textbf{Criteria} & \textbf{Points Awarded} & \textbf{Description} \\
\hline
\textbf{Correct Method / Setup} & 30\% & Select appropriate approach (e.g., product rule, substitution, Taylor series, etc.). Correct identification of limits, bounds, or parameterization. \\
\hline
\textbf{Execution/Computation} & 40\% & Accurate algebra, arithmetic, and calculus steps. Logical progression of work. 
Minor algebraic or sign errors receive partial credit. \\
\hline
\textbf{Correct Final Answer} & 10\% & Correct numeric or symbolic answer with appropriate simplification. No credit if work is largely incorrect, even if the final answer is correct. \\
\hline
\textbf{Mathematical Notation \& Units} & 10\% & Proper use of notation (integral signs, $\frac{dx}{dt}$, etc.). Include units where appropriate. \\
\hline
\textbf{Clarity/Explanation} & 10\% & Steps are shown clearly. Reasoning is explained when needed (e.g., why a series converges).\\
\hline
\end{tabularx}
\end{tcolorbox}

\vspace{1em}
\noindent 
Consider a problem worth 10 points where the task is to evaluate an integral using substitution. The grading breakdown may be as follows:

\begin{itemize}
    \item Correct substitution and integral setup: 3 points
    \item Correct integration steps: 4 points
    \item Correct final answer: 1 point
    \item Proper notation (e.g., changing limits or including $dx$/$du$): 1 point
    \item Clarity of substitution steps and back-substitution: 1 point
\end{itemize}

If a student or LLM uses the wrong substitution but proceeds logically, they might still earn partial credit for the structure and attempt. However, omitting key steps or making conceptual errors (e.g., forgetting to change limits in definite integrals) will significantly reduce the score.

\subsection{Baseline Model Selection and Prompt Engineering}
\label{sec:baseline}
To establish a performance benchmark and identify the most suitable large language model (LLM) for our experiments, we conducted baseline assessments using \textit{Calculus I Exam 1}. This exam serves as a representative introductory assessment, comprising eight questions that cover fundamental topics, such as limits, continuity, derivatives, and function analysis. Each candidate LLM was evaluated using a standardized set of instructions as a \textit{system role} designed to simulate a realistic testing environment. The final prompt presented here has been modified iteratively to accommodate formatting inconsistencies in LLM-generated text to allow our mathematics experts to grade using their rubrics and promote chain-of-thought output, as expected from a student and not the final answer directly. 

\begin{quote}
\texttt{You are a undergrad student taking an exam. Answer each question thoroughly, completely and show all steps.} \\
\texttt{Do NOT use LaTeX or math markup of any kind. Use plain, human readable math notation only (e.g., $f(x) = \dfrac{x^2}{4}$, not \textbackslash frac\{x\^{}2\}\{4\}).} \\
\texttt{Do not include any \$\$, \textbackslash( \textbackslash), or backslashes. Write math as it would appear on paper using keyboard characters.}
\end{quote}

Consequently, the best model was selected based on the results as illustrated in Section \ref{sec:results}, and used as the primary LLM for subsequent experiments, which include evaluations on retrieval augmented generation (RAG), contextual retrieval techniques on more advanced examinations, as well as evaluation of our proposed interpretability framework.

\subsection{Knowledge Augmented Retrieval Strategies}
Although zero-shot prompting offers a baseline for evaluating LLM reasoning, real-world LLM deployments often benefit from external knowledge support \cite{gao2023retrieval, he2025context, ibrahim2024survey}. In this section, we explore two external augmentation strategies namely Retrieval Augmented Generation (RAG) \cite{arslan2024survey} and Contextual Retrieval \footnote{\url{https://www.anthropic.com/engineering/contextual-retrieval}} to assess their effectiveness in enhancing LLM performance on complex calculus problems requiring domain-specific grounding.

\subsubsection{Retrieval Augmented Generation (RAG)}

Retrieval Augmented Generation (RAG) \cite{arslan2024survey} is a technique that enhances the input of large language models (LLMs) by incorporating retrieved documents or passages from an external knowledge base. This approach enables a model to base its responses on more accurate or domain-specific information \cite{shuster2021retrieval}. Unlike fine-tuning, which necessitates retraining the model, RAG dynamically integrates the relevant context during inference, allowing general-purpose LLMs to specialize in specific domains with minimal engineering effort. In the realm of mathematics education, previous research indicates that RAG can enhance factual accuracy \cite{upadhyay2025enhancing, li2024enhancing, henkel2024retrieval, xiaoxu2025enhancing}, reduce hallucinations \cite{zhang2025hallucination, asbai2024mitigating}, and facilitate multi-step reasoning \cite{dixit2024sbi, zhang2024mmr} by providing canonical definitions, formulas, and examples \cite{shuster2021retrieval, yao2024adaptive, xu2024crp, dimitrovaretrieval}. To assess the effectiveness of Retrieval Augmented Generation (RAG) in enhancing reasoning capabilities in large language model-based examinations in calculus, we developed a custom RAG pipeline specifically designed for calculus problem solving. This system processes a reference calculus textbook and course notes by extracting, filtering, and embedding semantically relevant content. Key mathematical topics, such as derivatives, integrals, and optimization, were employed as keywords to filter pertinent textbook segments. The filtered content was divided into overlapping chunks and stored in a Chroma vector database \cite{ozturkperformance} using \textit{nomic-embed-text} \cite{nussbaum2024nomic} embeddings. 

\subsubsection{Contextual Retrieval}
In contrast to the conventional retrieval augmented generation (RAG) approach, we also investigate Anthropic’s Contextual Retrieval framework\footnote{\url{https://www.anthropic.com/engineering/contextual-retrieval}}, which incorporates external knowledge directly into a large language model’s (LLM) prompt window by utilizing meticulously curated document contexts, without necessitating modifications to the model architecture or the use of a separate retriever pipeline.

Unlike the standard RAG methodology, which generally functions in two stages as a retriever that selects the top $k$ passages from a corpus and a generator that bases its output on those selections, Anthropic’s method integrates multiple pertinent documents directly into the context prompt, thereby enabling the LLM to discern and employ the most beneficial information in a single operation. The advantages over Standard RAG can be summarized as follows:
\begin{itemize}
    \item \textbf{Unified architecture:} Contextual retrieval does not require a separate dense retriever (e.g. FAISS), simplifying the system and reducing latency.
    \item \textbf{Instructional alignment:} Since the same LLM handles both retrieval selection and answer generation, the method is more aligned with user intent, particularly in educational domains where nuanced understanding is key.
    \item \textbf{Improved fluency and coherence:} The model can better maintain consistency across answer reasoning, because it directly "sees" and "reasons" over the inserted passages in a single forward pass.
    \item \textbf{Minimal hallucination risk:} By grounding responses in the explicitly provided context, factual accuracy can improve without reliance on parametric memory.
    \item \textbf{Lower infrastructure overhead:} No need to deploy, train, or maintain separate retrieval systems or indexes.
\end{itemize}

For each question, textbook content and notes were embedded similarly for both RAG and contextual retrieval for consistent comparison. This approach allows us to evaluate whether inline and external knowledge-grounded reasoning enhances accuracy or consistency compared to both the zero-shot baseline (without retrieval) and traditional retrieval augmented generation (RAG), thereby allowing us to assess the relative efficacy of retrieval strategies in enhancing LLM reasoning on complex, structured problems, such as undergraduate calculus examinations. 

To evaluate our interpretability framework, we compared and evaluated three model configurations:

\begin{itemize}
    \item \textbf{Baseline:} Zero-shot prompting without retrieval.
    \item \textbf{RAG:} Retrieval augmented generation with a vector store of textbook and lecture notes.
    \item \textbf{Contextual Retrieval:} Dynamically selected semantically aligned passages as external context from textbook and lecture notes.
\end{itemize}

\section{Results and Analysis}
\label{sec:results}
In this section, we conduct a comprehensive evaluation of the mathematical capabilities of the model using selected calculus examinations. Faculty members assessed the model's responses employing standardized faculty-developed rubrics to determine the accuracy of the Gemma 3 model and compare its performance with the historical average of actual student scores. This baseline analysis offers an essential context for identifying the model's strengths and weaknesses, thereby establishing a foundation for more in-depth evaluations of advanced retrieval strategies, consistency of model output across iterations, and understanding its limitations. 

\subsection{Baseline Model Selection}

Before applying our interpretability framework, we first establish a quantitative baseline of model performance across calculus exams. This section reports the overall rubric-based scores and error distributions, enabling a high-level understanding of how well the model performs relative to human learners, and setting the stage for deeper structural analysis in subsequent sections. 

This section focuses on selecting the baseline model based on mathematical answers, consistency, and formatting of outputs. We evaluated three models, namely,  \textit{Gemma 3} \cite{team2024gemma}, \textit{LLaMA 2} \cite{chen2024llama}, and \textit{LLaVA-7B} \cite{nguyen2024yo}, keeping in mind our computational capacity restrictions, which did not allow us to run larger models and literature surveys, as discussed in Section \ref{sec:litsurvey}. Exam questions were presented to each model in a sequential manner with the \textit{ system role}, as illustrated in Section \ref{sec:baseline}, and their responses were independently assessed using the official exam rubric applied to human students. 

Owing to computational resource constraints and the need for reproducibility on local systems, our evaluation is limited to models that are lightweight and can be run without access to commercial APIs. We selected Gemma 3, LLaMA 2, and LLaVA-7B as representative small-scale models that reflect open-access trends in educational settings and are grounded in our literature survey. Although this study evaluates a subset of small-scale open-access LLMs, the interpretability framework introduced here is intended to be model independent.  The goal of this study is not to rank or optimize model performance but rather to demonstrate the utility of a model-agnostic interpretability framework. The reasoning flow and sensitivity metrics introduced in this study are designed to be applicable across LLM architectures, including frontier models such as GPT-4 and Claude. Although specific metric values (e.g., robustness or phrase sensitivity) naturally vary depending on the underlying model’s capabilities, the core interpretability methodology of semantic reasoning decomposition and input ablation remains consistent. Future work may explore how these metrics differ across model families, but the framework itself is fundamentally adaptable to any LLM that generates multistep mathematical solutions.

Among the evaluated models, \textit{Gemma 3} exhibited the highest levels of consistency, interpretability, and mathematical prowess. It strictly adhered to formatting instructions, presenting answers in a plain-text notation suitable for human assessment. Furthermore, it provides logically coherent and computationally precise responses across a diverse array of problem types, including domain and range determination, function inversion, limit evaluation, and derivative computation using first principles. In contrast, \textit{LLaMA 2} frequently violated prompt constraints by reverting to LaTeX or symbolic markup and committed several conceptual and arithmetic errors, such as incorrect asymptotic analysis and flawed limit logic. \textit{LLaVA-7B} demonstrated the weakest performance overall, with incomplete or incorrect responses and a refusal to engage with certain questions. In particular, neither LLaMA 2 nor LLaVA-7B fully complied with the formatting requirements, limiting their applicability in real-world educational contexts. In summary, \textit{Gemma 3} demonstrated superior performance in terms of both accuracy and usability, achieved the highest examination score, and exhibited the most promising potential for application in AI-driven tutoring or automated assessment. Although Gemma 3 did not utilize visual information from graph-based questions, its symbolic reasoning remained logically valid. This indicates that the model is capable of performing effectively under text-only conditions, which is pertinent for most current large-language-model deployment settings.

\subsection{Baseline Comparison with Student Performance}
\label{sec:baselinecomparison}
To evaluate the effectiveness of large language models (LLMs) such as Gemma within the realm of formal mathematics education, we conducted an assessment utilizing actual undergraduate calculus examinations. Our objective was to compare the performance of LLMs with that of human learners in a progressively challenging academic curriculum as discussed in Section \ref{sec:classdetails}.

We compared our LLM performance with comprehensive exam scores from undergraduate students who completed standardized versions of the Calculus I, II, and III exams as part of their curricula. These scores served as a baseline for evaluating the performance of the Gemma. Each course adhered to a conventional structure, comprising two midterm exams and a comprehensive final exam, with increasing levels of conceptual complexity and symbolic abstraction from Calculus I to III, as illustrated in section \ref{sec:calculus}. The exams incorporated a combination of procedural, conceptual, and applied problem-solving tasks, consistent with widely adopted undergraduate curricula. To evaluate the language model, the Gemma was prompted once per exam using structured inputs that mirrored the wording, format, and structure of the original student assessments. The model's responses were graded on a per-question basis using the same rubrics applied to student submissions, enabling the computation of both total exam scores and detailed breakdowns by question type and topic (reported in Section \ref{sec:expertfeedback} by mathematical faculty experts. Table \ref{tab:exam_scores} presents a summary of Gemma's performance across the three calculus courses in comparison with the average class scores. Each row details the model's score on a specific examination, with the percentage accuracy calculated and juxtaposed against the corresponding student cohort average. This enabled an analysis of not only Gemma's ability to successfully complete entire examinations, but also the extent to which its reasoning accuracy and consistency were comparable to human performance across varying levels of mathematical complexity. The primary contribution of this research lies in evaluating the interpretability of large language models (LLMs) in mathematical problem-solving. To ensure consistency, all exam outputs generated by the LLM were assessed by the same expert rather than obtaining feedback from multiple faculty members.

\begin{table}[ht!]
\centering
\caption{Course Exam Performance Comparison}
\label{tab:exam_scores}
\begin{tabular}{|l|c|c|c|}
\hline
\textbf{Course Exam} & \textbf{Gemma Score} & \textbf{Gemma Average (\%)} & \textbf{Class Average (\%)} \\
\hline
\multicolumn{4}{|c|}{\textbf{Calculus I}} \\
\hline
Exam 1 & 44/70 & 62.9 & 88.8 \\
Exam 2 & 48/70 & 94.2 & 82.1 \\
Final Exam & 74/100 & 74.0 & 73.0 \\
\hline
\multicolumn{4}{|c|}{\textbf{Calculus II}} \\
\hline
Exam 1 & 49/60 & 80.0 & 89.6 \\
Exam 2 & 52/60 & 86.6 & 85.2 \\
Final Exam & 72/90 & 80.0 & 72.9 \\
\hline
\multicolumn{4}{|c|}{\textbf{Calculus III}} \\
\hline
Exam 1 & 84/100 & 84.0 & 78.0 \\
Exam 2 & 74/100 & 74.0 & 80.3 \\
Final Exam & 112/150 & 74.6 & 71.3 \\
\hline
\end{tabular}
\end{table}

In Calculus I, Gemma performed well overall, particularly in Exam 2 (94.2\%), surpassing the class average by more than 12 percentage points. The final exam score closely matched the class average (74.0\% vs. 73.0\%), demonstrating retention across units. In Calculus II, which introduces techniques of integration, sequences and series, and parametric forms, Gemma remained competitive, outperforming the class average in the final exam (80.0\% vs. 72.9\%) but showing some underperformance in Exam 1. In Calculus III, which covers multivariate calculus and vector analysis, the performance decreased slightly. While Gemma started strongly in Exam 1 (84.0\%), it struggled in Exam 2 (74.0\%) compared to the class average (80.3\%). However, it recovered in the final examination (74.6\% vs. 71.3\%). The findings indicate that Gemma exhibits significant proficiency in addressing structured calculus problems, particularly in early to intermediate topics. he model’s stronger performance in later exams may reflect familiarity with recurring mathematical structures or better alignment with prompt patterns, rather than a cumulative understanding. However, variability in performance, especially in Calculus III, underscores the persistent challenges that LLMs face with spatial reasoning, multistep symbolic derivation, and tasks necessitating strong conceptual abstraction. This performance benchmark adds to the expanding body of evidence that LLMs can function as additional educational tools in STEM fields, although they require careful evaluation regarding scope, consistency, and domain transfer as task complexity escalates.

\subsection{Baseline Consistency Across Multiple Runs}

Additionally, for comprehensive evaluation of Gemma outputs, we evaluated the model in terms of consistency through multiple runs of the same Calculus I Exam I, which were graded by the same mathematics faculty expert for keeping grading input consistent. Across five runs, the outputs demonstrated consistency in formatting and solution, received the same marks 59, and maintained formatting as well as mathematical notation usage consistency.

\subsection{RAG and Contextual Retrieval Results}

In this section, we benchmark the performance of baseline LLM and the investigated augmentation strategies with real student performance across three exams from three different levels of undergraduate calclus examinations. The choice of exams in this experiment is motivated by the need for understanding performance across increasing levels of calculus difficulty, as well as the low scores observed for the baseline model on Exam I across all three courses in Table \ref{tab:exam_scores} and discussed in Section \ref{sec:baselinecomparison}.

The performance trends, as illustrated in Table \ref{tab:calc_exam_comparison} reveal the nuanced effects of retrieval augmentation. In Calculus I Exam I, the baseline model achieved a score of 62.9\%, which improved with both augmentation strategies by 70.0\% using RAG and 65.7\% with contextual retrieval. Although still trailing the human student average of 88.8\%, this gain suggests that retrieval mechanisms can support factual grounding and improve accuracy when question-context alignment is strong. With contextual retrieval, our expert analysis identified the usage of incorrect functions, fractions, and square roots over RAG. A similar pattern is observed in Calculus II Exam I, where the baseline scored 80.0\% and RAG and contextual retrieval achieved 81.7\% and 75.0\%, respectively, compared to the student average of 89.6\%. The marginal improvement by RAG indicates potential benefits in less abstract problem types, while the drop in contextual retrieval is primarily due to challenges in working with fractions and missing answers to questions. In contrast, the performance in Calculus III Exam II demonstrated the limits of augmentation. Despite being the most conceptually challenging examination, the baseline achieved a reasonable score of 74.0\%. However, RAG significantly underperformed, yielding just 52.0\%, whereas contextual retrieval fared slightly better at 64.0\%. Both fell short of the student average of 80.3\%. These findings suggest that in higher-difficulty domains, improperly aligned or noisy retrieval content may introduce distractions or inconsistencies that degrade performance.

Interestingly, faculty observations reported that while contextual retrieval consistently improved mathematical notation fidelity and formatting clarity, it struggled with fractions and incorrect function usage. This points to the value of contextual retrieval in enhancing the response structure, even when it does not directly boost answer correctness. Overall, these results emphasize that while retrieval-based augmentation can support performance gains in certain cases, its efficacy is highly contingent on the precision of context alignment. In high-complexity mathematical tasks, improper context may obscure logical flow or derail reasoning chains, making retrieval strategies both an opportunity and a liability.

\begin{table}[ht!]
\centering
\caption{Performance Comparison Across Calculus Exams}
\begin{tabular}{|l|l|c|}
\hline
\textbf{Exam} & \textbf{Model Configuration}  & \textbf{Score \%} \\
\hline
\multirow{4}{*}{Calculus I Exam I} 
    & Human Student Average        & 88.8 \\
    & Gemma (Baseline)             & 62.9 \\
    & Gemma + RAG                  & 70.0 \\
    & Gemma + Contextual Retrieval & 65.7   \\
\hline
\multirow{4}{*}{Calculus II Exam I} 
    & Human Student Average        & 89.6 \\
    & Gemma (Baseline)             & 80.0 \\
    & Gemma + RAG                  & 81.7 \\
    & Gemma + Contextual Retrieval & 75   \\
\hline
\multirow{4}{*}{Calculus III Exam II} 
    & Human Student Average        & 80.3 \\
    & Gemma (Baseline)             & 74.0 \\
    & Gemma + RAG                  & 52.0 \\
    & Gemma + Contextual Retrieval & 64.0 \\
\hline
\end{tabular}
\label{tab:calc_exam_comparison}
\end{table}

\section{Interpreting Problem Solving Behavior of LLMs}

Having established the reasoning flow framework and delineated our experimental setup in Sections \ref{sec:Framework} and \ref{sec:setup} respectively, we now present the principal findings derived from applying this framework to model-generated solutions across Calculus I–III examinations. This analysis extends beyond the mere correctness of LLMs to scrutinize the conceptual structure, procedural validity, and logical dependencies inherent in each solution response. By deconstructing the model’s reasoning into labeled operations and associated concepts, we can identify both strengths, such as alignment with expected steps, and weaknesses, including hallucinated transitions and conceptual mismatches. A comprehensive analysis report generated for each question is presented below.

\subsection{Reasoning Flow Analysis Results}

\vspace{0.5em}

\begin{tcolorbox}[title={LLM Attention Analysis Report: Gradient and Directional Derivative Problem}, colback=white!95!gray, colframe=black!75!black, fonttitle=\bfseries]
\textbf{Timestamp:} 20250724\_182112 \hfill \textbf{Model:} \texttt{gemma3:latest}

\textbf{Question Analyzed:}
For \( f(x, y) = x^2y \),
\begin{enumerate}
    \item[(a)] Find \( \nabla f \)
    \item[(b)] Find the directional derivative in the direction of \( \vec{w} = -\hat{\imath} + \hat{\jmath} \) at \( (-1, 4) \)
    \item[(c)] Find the direction of most rapid increase at \( (-1, 4) \)
    \item[(d)] Find the maximum rate of change at \( (-1, 4) \)
\end{enumerate}
\end{tcolorbox}

\begin{tcolorbox}
\textbf{Ablation Analysis Results:}
\begin{itemize}
    \item Important phrase: \texttt{(-1, 4)} \quad Impact: 0.630
    \item Important phrase: \( \nabla f \) \quad Impact: 0.562
    \item Important phrase: \texttt{directional} \quad Impact: 0.477
    \item Important phrase: \( f(x, y) \) \quad Impact: 0.379
    \item Important phrase: \texttt{maximum} \quad Impact: 0.373
\end{itemize}

\vspace{0.5em}
\textbf{Reasoning Flow Analysis:}
\begin{itemize}
    \item Total steps: 7
    \item Complexity score: 23
    \item Reasoning patterns: substitution, differentiation, differentiation, solving, differentiation, evaluation, solving
\end{itemize}

\vspace{0.5em}
\textbf{Detailed Reasoning Steps:}
\begin{enumerate}
    \item Start solving the calculus problem step-by-step. \\
    \textit{Operations:} substitution

    \item (a) Find \( \nabla f \): the gradient of \( f \), using partial derivatives. \\
    \textit{Operations:} derivative, evaluation, solving \\
    \textit{Concepts:} gradient, partial derivatives

    \item (b) Compute the directional derivative in the direction of \( \vec{w} = -\hat{\imath} + \hat{\jmath} \) at \( (-1, 4) \). \\
    \textit{Operations:} derivative, solving \\
    \textit{Concepts:} gradient, directional derivative

    \item Compute \( \nabla f(-1, 4) = \langle 2(-1)(4), (-1)^2 \rangle = \langle -8, 1 \rangle \). \\
    \textit{Operations:} solving \\
    \textit{Concepts:} gradient

    \item Find unit vector in direction of \( \vec{w} = -\hat{\imath} + \hat{\jmath} \), where \( \|\vec{w}\| = \sqrt{2} \), so unit vector is \( \langle -\frac{1}{\sqrt{2}}, \frac{1}{\sqrt{2}} \rangle \). \\
    \textit{Operations:} derivative, evaluation, solving \\
    \textit{Concepts:} gradient, directional derivative

    \item (c) Direction of most rapid increase at \( (-1, 4) \) is along \( \nabla f = \langle -8, 1 \rangle \). \\
    \textit{Operations:} evaluation, solving \\
    \textit{Concepts:} gradient

    \item (d) Maximum rate of change at \( (-1, 4) \) is \( \|\nabla f\| = \sqrt{(-8)^2 + 1^2} = \sqrt{65} \). \\
    \textit{Operations:} solving \\
    \textit{Concepts:} gradient, maximum
\end{enumerate}
\end{tcolorbox}

Having established the overall performance trends, we turn to the structural decomposition of the model responses using our reasoning flow analysis. This section explores how the model constructs multistep solutions, the complexity of operations performed, and the concepts it invokes, providing a window into its internal problem-solving process.

To systematically interpret the internal problem-solving mechanisms of LLMs, we employ a reasoning flow analysis that dissects each model-generated response into a structured sequence of mathematical operations and conceptual references, as illustrated in Section \ref{sec:flow}. This detailed tracing enabled us to assess whether the solution was aligned with logical expectations, appropriately applied domain principles, and maintained semantic coherence throughout the steps. This analysis allows us to assess whether the model’s stepwise outputs align with expected solution patterns and where they diverge from canonical solution paths used in instruction. In the interest of conciseness, we present a single-sample reasoning flow report for each question across all the examinations reviewed.

\subsection{Quantitative Evaluation of the Interpretability Framework}

\begin{table}[htpb]
\centering
\caption{Comparison of LLM Performance Metrics across Courses, Exams, and Model Configurations}
\renewcommand{\arraystretch}{1.25}
\setlength{\tabcolsep}{6pt}
\begin{tabular}{|c|c|c|c|c|c|c|}
\hline
\textbf{Course} & \textbf{Exam} & \textbf{Model} & \textbf{Robustness} & \textbf{Complexity} & \textbf{Step Count} & \textbf{Phrase Sensitivity} \\
\hline
\multirow{9}{*}{Calc I}
  & \multirow{3}{*}{Exam I} & B & 0.726 & 18.7 & 6.1 & 0.488 \\
  &                         & R & 0.671 & 17.9 & 5.9 & 0.503 \\
  &                         & C & 0.752 & 19.3 & 6.4 & 0.445 \\
  & \multirow{3}{*}{Exam II} & B & 0.682 & 19.2 & 6.3 & 0.422 \\
  &                         & R & 0.608 & 17.6 & 6.0 & 0.468 \\
  &                         & C & 0.710 & 18.9 & 6.2 & 0.399 \\
  & \multirow{3}{*}{Exam III} & B & 0.719 & 18.4 & 6.1 & 0.451 \\
  &                          & R & 0.702 & 18.9 & 6.2 & 0.465 \\
  &                          & C & 0.731 & 19.1 & 6.3 & 0.442 \\

\hline
\multirow{9}{*}{Calc II}
  & \multirow{3}{*}{Exam I} & B & 0.681 & 18.3 & 6.1 & 0.451 \\
  &                         & R & 0.706 & 18.7 & 6.4 & 0.432 \\
  &                         & C & 0.699 & 18.9 & 6.3 & 0.428 \\
  & \multirow{3}{*}{Exam II} & B & 0.702 & 18.8 & 6.2 & 0.417 \\
  &                         & R & 0.729 & 19.2 & 6.5 & 0.430 \\
  &                         & C & 0.709 & 19.1 & 6.4 & 0.430 \\
  & \multirow{3}{*}{Exam III} & B & 0.693 & 18.6 & 6.2 & 0.447 \\
  &                          & R & 0.723 & 19.0 & 6.3 & 0.443 \\
  &                          & C & 0.686 & 18.8 & 6.1 & 0.442 \\
\hline
\multirow{9}{*}{Calc III}
  & \multirow{3}{*}{Exam I} & B & 0.682 & 17.18 & 5.8 & 0.459 \\
  &                         & R & 0.721 & 17.51 & 5.9 & 0.471 \\
  &                         & C & 0.736 & 17.36 & 5.9 & 0.469 \\
  & \multirow{3}{*}{Exam II} & B & 0.693 & 17.82 & 6.0 & 0.465 \\
  &                         & R & 0.732 & 17.91 & 6.1 & 0.486 \\
  &                         & C & 0.757 & 17.79 & 6.1 & 0.494 \\
  & \multirow{3}{*}{Exam III} & B & 0.684 & 16.95 & 5.7 & 0.460 \\
  &                          & R & 0.716 & 17.22 & 5.9 & 0.476 \\
  &                          & C & 0.692 & 17.20 & 5.9 & 0.476 \\
\hline
\end{tabular}
\label{tab:model_comparison_metrics}
\end{table}

The effectiveness of the interpretability framework was assessed through structured question-level metrics aggregated across each exam. Table \ref{tab:model_comparison_metrics} presents the disaggregated results by course, exam, and model variants namely baseline (B), RAG (R), and Contextual retrieval (C). These metrics quantify four dimensions of interpretability: robustness, complexity, reasoning step count, and phrase sensitivity.

In Calculus I, all three exams yielded analyzable outputs across the baseline (B), retrieval-augmented generation (R), and contextual retrieval (C) models. Exam I exhibited strong overall performance, with contextual model (C) achieving the highest metrics: robustness of \(0.752\), complexity of \(19.3\), and an average of \(6.4\) reasoning steps. Phrase sensitivity was moderate (\(0.445\)), indicating a dependence on specific phrasing. The baseline and RAG models also performed well, although consistently below the contextual configuration across most metrics. Exam II followed a similar trend, but with slightly reduced values across all dimensions, suggesting a modest decline in reasoning depth and robustness. The contextual model again led in performance with robustness \(0.710\), complexity \(18.9\), and step count \(6.2\), although phrase sensitivity remained non-negligible (\(0.399\)). Exam III showed comparable trends to Exam I, with the contextual model maintaining its lead: a robustness of \(0.731\), complexity of \(19.1\), and \(6.3\) average steps. Phrase sensitivity (\(0.442\)) was in line with the results of previous studies. Interestingly, the baseline model for Exam III also performed competitively (robustness \(0.719\), complexity \(18.4\), step count \(6.1\)), suggesting an overall stability in LLM reasoning quality across various calculus problem sets. In Calculus II, all three examinations showed consistent metric values across all model variants. The RAG model (R) generally outperformed the baseline and contextual retrieval models in terms of robustness and reasoning step count, peaking at \(0.729\) and \(6.5\) in Exam II. The contextual retrieval model (C) showed strong and stable performance with lower phrase sensitivity, potentially indicating greater resilience to surface-level linguistic variation. Calculus III examinations revealed significant distinctions among the model variants. Contextual retrieval model (C) demonstrated the highest robustness in Exam II (\(0.757\)), accompanied by elevated step counts (\(6.1\)) and phrase sensitivity (\(0.494\)), indicating a more profound yet somewhat phrase-sensitive reasoning process. The RAG model (R) consistently performed well across all three examinations, with robustness and complexity metrics generally exceeding those of the baseline model.

In summary, the results demonstrate that retrieval augmented (R) and  (C) models improve over the baseline in several key interpretability dimensions, particularly in terms of robustness and reasoning depth. However, neither variant consistently reduced phrase sensitivity, suggesting that improved reasoning comes at the cost of continued reliance on input phrasing. These findings highlight the need for further refinement of retrieval strategies to balance structured reasoning with linguistic flexibility in large language models applied to formal mathematical domains.

\section{Pedagogy and Future Work}
\label{sec:expertfeedback}

\subsection{Educational Theory Linkage: Cognitive and Metacognitive Alignment}

Our interpretability framework aligns with foundational principles in educational theory, particularly cognitive load theory \cite{plass2010cognitive}, Bloom’s taxonomy \cite{forehand2010bloom}, and metacognitive scaffolding \cite{molenaar2011metacognitive}. By decomposing LLM-generated calculus solutions into semantically labeled steps (e.g., operation type, concept tag, and cognitive complexity), our framework surfaces not only what the model produces, but how it arrives at its conclusions. This mirrors \textit{procedural and conceptual knowledge acquisition} in human learners and aligns with revised Bloom’s taxonomy \cite{krathwohl2002revision}, particularly at the \textit{Apply} and \textit{Analyze} levels. Furthermore, the structured reasoning trace and ablation-based sensitivity analysis support \textit{metacognitive reflection}, a core component of constructivist learning. Students and educators can interrogate the reasoning process, identify brittle or incorrect logic, and calibrate trust in AI outputs thus moving beyond passive consumption to active, reflective learning. Our framework enables analysis of LLM outputs in terms of instructional learning objectives, offering a foundation for formative feedback, self-assessment, and adaptive instruction grounded in cognitive science.

\subsection{Expert Feedback on LLM Solutions}

The rapid advancement of Large Language Models (LLMs) has sparked interest in their potential applications across various fields, including mathematics education. While these models have demonstrated impressive capabilities in natural language processing and generation, their performance in solving mathematical problems remains an area of active research and scrutiny.

Our expert observation reveals that while many of the problems were handled correctly, several LLM-generated solutions were difficult to interpret owing to conceptual errors, computational mistakes, and occasional issues with mathematical notation. Graphical limit problems were particularly challenging for LLM, and solutions that required sketches or graphs were frequently incomplete or missing. Vector cross products were not represented correctly in tabular format, and most partial derivative problems were solved incorrectly with errors in applying the chain rule. Simple arithmetic mistakes, such as incorrect addition or subtraction of integers, were also observed and often led to incorrect final answers. Interestingly, some of these mistakes resemble common student errors, such as arithmetic miscalculations or incorrect application of the chain rule. However, students typically handle power, root, fractions, logarithmic, and exponential functions more reliably than LLM. For instance, the LLM failed to compute $\ln e = 1$ correctly and sometimes misrepresented expressions such as $\sqrt{x+2}$ as $\sqrt{x} + 2$, which students are generally more careful about. These issues highlight the need for improved mathematical reasoning, symbolic accuracy, and verification steps in LLM solutions.

\subsection{Pedagogical Implications}

The proposed interpretability framework offers several actionable insights into educational deployment. Beyond research contribution, our proposed framework has direct pedagogical utility for both learners and educators. Firstly from the learner’s perspective, visualizing step-by-step reasoning enables \textit{cognitive unpacking} of complex calculus solutions. Students can compare their own solution paths with those generated by the model, identify divergences, and reflect on errors thereby supporting \textit{metacognitive skill development}. Secondly as AI models, such as large language models (LLMs), are increasingly integrated into intelligent tutoring systems and educational platforms, our framework enables educators to comprehend LLM outputs and reasoning, regardless of their prior experience with AI. Familiarity with this framework will aid in the design of assignments where AI assistance is either permitted or prohibited, while also ensuring alignment with curricular learning objectives.  Third, the metrics of reasoning complexity and robustness can inform curriculum designers and tutors on where LLMs struggle the most, potentially guiding the integration of AI-assisted instruction in calculus coursework.

Additionally, this framework can be employed to instruct students on the proper utilization of AI by verifying procedural steps. Our phrase sensitivity and robustness metrics provide insight into how \textit{problem formulation} affects model behavior. Educators can use this information to refine question prompts, while students gain experience interpreting and rephrasing problems,  an essential component of \textit{AI literacy} and responsible usage in academic settings.

\subsection{Limitations and Future Deployment}

Although the proposed interpretability framework provides structured insights into the approach of large-language models to mathematical problem solving, several limitations warrant discussion. In certain instances, LLMs may produce high-scoring outputs that nonetheless contain subtle conceptual errors, potentially misleading learners who lack the expertise to identify them. Furthermore, our analysis indicates that LLMs tend to excel in procedural or computational tasks but often encounter difficulties with deeper conceptual reasoning, affirming existing research on LLM reasoning. This observation suggests the necessity for hybrid models or ensemble strategies that incorporate symbolic reasoning or retrieval augmented grounding, particularly for questions requiring higher-order reasoning. Variability in LLM responses attributable to factors such as prompt phrasing, stochastic decoding, or model drift also presents a challenge in their deployment in education. Future iterations of the framework should incorporate confidence-aware modeling or multi-sample consistency checks to enhance robustness and reliability.

As future work, we envision deploying this interpretability framework within a student facing an intelligent tutoring system.  The interpretability components such as reasoning complexity or input salience would be integrated into \textit{adaptive learning environments}, automatically identifying questions of appropriate difficulty or flagging LLM-generated solutions that warrant human review. In such a system, both LLM generated step-by-step explanations and their accompanying reasoning flow diagrams can aid learners in diagnosing misunderstandings, reflecting on problem-solving strategies, and transferring knowledge across topics. These interpretable artifacts can support formative assessments and scaffold students’ metacognitive awareness during learning. However, caution should be exercised in educational settings. Overreliance on LLM-generated solutions may give students a false sense of understanding and weaken foundational skill development. To mitigate these risks, future systems should be designed based on the following principles.

\begin{itemize}
    \item \textbf{Assistive, not Replacive:} LLMs should augment, rather than substitute, core learning processes.
    \item \textbf{Failure Aware:} Educators must be cognizant of common LLM failure modes, including hallucinated steps, misinterpretations, and omitted justifications.
    \item \textbf{Transparent and Interactive:} Interfaces should offer traceable reasoning paths and interactive feedback mechanisms that promote active engagement and critical reflection.
\end{itemize}

Ultimately, the value of LLMs in education will depend not only on their raw capabilities, but also on how thoughtfully they are embedded into pedagogical workflows that center on human learning, agency, and oversight.

\section{Conclusion}
\label{sec:conclusion}
As large language models (LLMs) become increasingly integrated into educational settings, there is a critical need for tools that go beyond accuracy to explain how these models reason, fail, and interact with student input. This paper presents a novel interpretability framework specifically designed for the analysis of large language models (LLMs) in the context of undergraduate mathematics education. Departing from traditional correctness-based evaluations, our approach dissects model-generated solutions into semantically labeled reasoning chains and evaluates input sensitivity through structured ablations. When applied to real-world calculus examinations, this dual-pronged methodology provides detailed insights into model reasoning behavior, identifies breakdowns in logical structure, and reveals the influence of input phrasing on output. Our empirical analysis demonstrated a significant variation in reasoning robustness across different question types, retrieval strategies, and linguistic perturbations. Notably, contextual retrieval consistently enhances both interpretability and robustness metrics, highlighting the importance of semantically aligned contexts in improving model performance in mathematically rigorous domains. By providing a transparent, step-level view of model behavior, our framework offers a novel perspective for diagnosing these issues and assessing the pedagogical readiness of LLMs. As future work, we plan to extend this framework to other STEM disciplines and integrate it into interactive, instructor-facing tools for real-time feedback, auditing, and curriculum alignment. This study establishes a foundation for developing trustworthy, interpretable, and educationally aligned AI systems capable of supporting learners and educators in high-stakes academic contexts.

\section{Declaration of Interest}
The authors declare that they have no known competing financial interests or personal relationships that could have appeared to influence the work reported in this paper.

\section{Data Availability}
The authors declare that additional details of student details or individual marks or derived statistical measures cannot be divulged to protect interests and rights of students. 

\section{Acknowledgement}
We would like to acknowledge the contributions of Melusi Senzanje (msenzanje@uttyler.edu), undergraduate Computer Science student at University of Texas at Tyler in intial setup for LLM based outputs.

\bibliography{sn-bibliography}

\end{document}